# Interfaces of high efficient kesterite $Cu_2ZnSnS(e)_4$ thin film solar cells[*]

Shoushuai Gao(高守帅)[1], Zhenwu Jiang(姜振武)[1], Li Wu(武莉)[2], Jianping Ao(敖建平)[1][†], Yu Zeng(曾玉)[1], Yun Sun(孙云)[1], and Yi Zhang(张毅)[1][‡]

1. *Institute of Photoelectronic Thin Film Devices and Technology,Tianjin Key Laboratory of Photoelectronic Thin Film Devices and Technology, Nankai University, Tianjin 300071, PR China*
2. *The MOE Key Laboratory of Weak-Light Nonlinear Photonics, School of Physics, Nankai University, Tianjin 300071, PR China*

## Abstract

$Cu_2ZnSnS(e)_4$ (CZTS(e)) solar cells have attracted much attention due to the elemental abundance and the non-toxicity. However, the record efficiency of 12.6% for $Cu_2ZnSn(S,Se)_4$ (CZTSSe) solar cells is much lower than that of $Cu(In,Ga)Se_2$ (CIGS) solar cells. One crucial reason is the recombination at interfaces. In recent years, large amount investigations have been done to analyze the interfacial problems and improve the interfacial properties via a variety of methods. This paper gives a review of progresses on interfaces of CZTS(e) solar cells, including: (1) the band alignment optimization at buffer/CZTS(e) interface, (2) tailoring the thickness of $MoS(e)_2$ interfacial layers between CZTS(e) absorber and Mo back contact, (3) the passivation of rear interface, (4) the passivation of front interface, and (5) the etching of secondary phases.

## 1. Introduction

[*] Project supported by the National Science Foundation of China (51572132, 51372121, 61674082), Tianjin Natural Science Foundation of Key Project (16JCZDJC30700), YangFan Innovative and Entrepreneurial Research Team Project (2014YT02N037), and 111 Project (B16027).

[†] Corresponding author. E-mail: aojp@nankai.edu.cn
[‡] Corresponding author. E-mail: yizhang@nankai.edu.cn

Cu(In,Ga)Se$_2$ (CIGS) solar cells are one of the most promising thin film photovoltaic devices with the record efficiency of 22.6%.[1] However, the reserves of indium (In)and gallium (Ga) in the earth is rare and the demand of In in display industry is huge, thus the maximum power production capacity of CIGS solar cells will be limited to ~ 100 GW per year.[2-3] Kesterite structure Cu$_2$ZnSnS(e)$_4$ (CZTS(e)) is derived from chalcopyrite structure CIGS with the more earth-abundant and low-cost elements Zn and Sn to replace the rare elements In and Ga.[4-5] Thus, CZTS(e) solar cells will be much cheaper. Besides, the optoelectronic properties of CZTS(e), such as appropriate band gap (1.0-1.5 eV varying with S/Se ratio) and high absorption coefficient (~10$^4$ cm$^{-1}$), make it ideal for use as an absorber material.[6-7]

In the past several years, CZTS(e) solar cells have attracted much attention, and the world record efficiency of Cu$_2$ZnSn(S,Se)$_4$ (CZTSSe) solar cells has achieved to 12.6% by Mitzi et al. in 2013.[8] But it still has a big gap with the Shockley-Queisser (S-Q) limit of the conversion efficiency of 32.8% for a CZTSSe solar cell.[9] Comparing the device performances of record CZTSSe and CIGS solar cells, as shown in Table 1, the major performance gap between CZTSSe solar cells and CIGS solar cells is the large open circuit voltage ($V_{OC}$) deficit, expressed as $E_g$/q- $V_{OC}$, where $E_g$ is the band gap and q is the elemental charge. Therefore, the $V_{OC}$ deficit is the key limiting factor for high efficient CZTS(e) solar cells.

To identify the reasons of big $V_{OC}$ deficit for CZTSSe solar cells, Gunawan et al. studied the temperature dependence of $V_{OC}$ for CIGS and CZTSSe solar cells to obtain the dominant recombination mechanism, as shown in Fig. 1.[10] The relationship between $V_{OC}$ and T can be expressed as[11]

$$V_{OC} = \frac{E_A}{q} - \frac{AkT}{q} \ln(\frac{J_{00}}{J_L}) \qquad (1)$$

where $E_A$, q, A, k, $J_{00}$, and $J_L$ are the activation energy of the dominant recombination mechanism, elemental charge, diode ideality factor, Boltzmann constant, reverse saturation current prefactor, and photo-current, respectively. Assuming A, $J_{00}$, and $J_L$ to be temperature independent, the $V_{OC}$ almost increase linearly with the decrease of T and the intercept of $V_{OC}$ vs. T plot at 0 K yields the activation energy $E_A$ of the dominant recombination mechanism.[11] As shown in Fig. 1, the activation energy $E_A$ of CIGS solar cell is almost the same as the band gap of the absorber, which indicates the dominant recombination process in CIGS solar cells occurs in the space charge region or bulk of the absorber.[12] However, the activation energy $E_A$ of both CZTSSe solar cells are significantly lower than their corresponding band gap, which is usually attributed to dominant recombination process at interface.[12-13] The interface recombination can be separated into three main categories: (1) A cliff-type band alignment at buffer/absorber interface, where the conduction band edge of the buffer layer is lower than that of the absorber, increasing recombination at interface; (2) Secondary phases at the interface trapping minority charge carriers or producing shunt paths; (3) A non-Ohmic back contact because of the over thick MoS(e)$_2$ between CZTS(e) and Mo interface, increasing the series resistance. This paper review the progress in the interface, including band alignment optimization, tailoring the thickness of MoS(e)$_2$ interfacial layer, interface passivation, and removal of secondary phases to show the perspective of how to improve the performance of kesterite solar cells.

## 2. Band alignment optimization at buffer/CZTS(e) interface

The band alignment at buffer/absorber interface is crucial for high efficient solar cells. There are two kinds of conduction band offset (CBO): cliff like and spike like band alignment, as shown in Fig. 2. For the cliff like band alignment, where the conduction band edge of buffer layer is below than that of absorber layer, the potential difference that can be generated between the quasi-Fermi levels of buffer layer and absorber layer under illumination will be reduced.[14] Besides, the cliff barrier will block the flow of injection electrons under forward bias, causing the increase of recombination between majority carriers via defects at buffer/absorber interface.[14-15] As a result, the $V_{OC}$ and $FF$ will be reduced. However, the spike like band alignment, where the conduction band edge of buffer layer is above that of absorber layer, is less detrimental to $V_{OC}$, and $V_{OC}$ is nearly constant despite the increase of CBO.[14-15] Besides, the simulation suggested that the $J_{SC}$ is nearly constant when the CBO is below 0.4 eV.[15] However, the photo-generated electrons cannot cross over the spike barrier when the CBO is too large (> 0.4 eV), leading to the decrease of $J_{SC}$ abruptly.[15] Therefore, a well band alignment at buffer/absorber interface is crucial for high efficient solar cells and the optimal range of CBO is 0 eV to 0.4 eV.[14-15]

### 2.1. CdS/CZTS(e) interface

Like CIGS solar cells, CdS is generally used as buffer layer in CZTS(e) solar cells. The champion CZTSSe solar cell with efficiency of 12.6% also used CdS as buffer layer.[8] To explore the reasons why the $V_{OC}$-deficit of CZTSSe solar cells is so large and further improve the efficiency, it is important to figure out the band alignment between CdS buffer and CZTSSe absorber. Table 2 shows the summary of CBO at CdS/CZTS or CdS/CZTSe interface. In 2011,

Bär et al. from Helmholtz-Zentrum Berlin für Materialien und Energie GmbH (HZB) studied the band alignment of CdS/CZTS heterojunction by means of X-ray photoelectron spectroscopy (XPS), ultraviolet photoelectron spectroscopy (UPS), and inverse photoelectron spectroscopy (IPES) spectra.[16] They found that the conduction band offset (CBO) and valence band offset (VBO) at CdS/CZTS interface is -0.33 eV and -1.2 eV, respectively, which means the CBO is cliff like. However, in the same year, Haight et al. from IBM also studied the band alignment at the interface between CdS buffer and $Cu_2ZnSn(S_xSe_{1-x})_4$ (CZTSSe) absorber as a function of S/(S+Se) ratio by UPS.[7] Their study indicated that the valence bandminimum (VBM) moves down with the increase of S/(S+Se) ratio, whereas, the conduction band minimum (CBM) is weakly dependent on S content. As a consequence, the VBO at CdS/CZTSSe interface decreases from 0.97 eV to 0.54 eV with the increase of S content, while the CBO is in the range of 0.4-0.5 eV, that is spike like and a little bit bigger than the optimal range (0-0.4 eV).[7] Afterwards, Tajima et al. measured the VBO at CdS/CZTS interface using X-ray photoelectron spectroscopy (HAXPES) and found that the value of VBO is ca. 1.0 eV, then the CBO were deduced as 0.0 eV.[17] Li et al. studied the CBO at CdS/CZTS and CdS/CZTSe interface and found that they are -0.06 and +0.34 eV, respectively.[18-19] Santoni et al. investigated the VBO at CdS/CZTS interface by XPS, and then the CBO were calculated, which is -0.3 eV.[20] Recently, Udaka et al. also investigated the band alignment at the interface between CdS buffer and $Cu_2ZnSn(S_xSe_{1-x})_4$ (CZTSSe) absorber with x=0, 0.3, and 1 by means of XPS, UPS, and IPES.[21] On contrary to Haight et al.,[7] they found that the CBM of CZTSSe obviously increases with the S-substitution of Se, and the CBM of CZTS is 0.40-0.45 eV higher than that of CZTSe, while the VBM only moves down 0.07 eV.[21] As a result, the conduction band offset (CBO) at CdS/CZTSSe interface decreases from +0.55 eV to -0.14 eV with the S-substitution of Se.[21] The CBO of +0.55 eV at Cd/CZTSe (x=0) interface is a rather high spike, which will block the transportation of photo-generated electrons and lead to a serious current-loss. Whereas, the negative CBO of -0.14 eV at CdS/CZTS (x=1) interface will cause serious voltage-loss. It is indicated that the CBO can be optimized by modifying the S/(S+Se) ratio, and the optimal ratio is in the range of 0.3-0.4. The conversion efficiencies of $Cu_2ZnSn(S_xSe_{1-x})_4$ solar cells with x= 0, 0.3, and 1are 7.2%, 11.4%, and 7%, respectively.[21] The CBO are also calculated using first principles. Bao et al. indicated that there are two inequivalent interfaces between CdS and CZTS, Cu-Zn layer or Cu-Sn layer adjacent to interface, respectively.[22] And the corresponding CBO calculated by first principles are -0.2 eV and -0.7 eV, respectively. If the CBO is averaged, then it is -0.3 eV. Palsgaard et al. calculated the CBO at CdS/CZTSe interface using first principles, which is spike like +0.6 eV.[23] The different results of CBO for the studies may be attributed to the difficulty in getting the accurate values of band bending near the surface, the different process conditions of absorber and buffer layer, and the different composition of the absorber layerespecially at the interface.[14] Thus, the results are inconclusive and more further studies are essential.

Overall, the CBO between CdS buffer and CZTSor CZTSe absorber is not optimal.There are following several methods to optimize the conduction band alignment at the buffer/absorber interface. As indicated in the above works, first, the CBO can be optimized by modifying the S/(S+Se) ratio.The optimal S/(S+Se) ratio is in the range of 0.3-0.4.[21] Udaka et al. indicated that the conversion efficiencies increased from 7.2% for CZTSe and 7% for CZTS to 11.4% for $Cu_2ZnSn(S_{0.3}Se_{0.7})_4$ solar cells because of the optimization of CBO.[21]

Besides, modifying the chemical bath depsition (CBD) process of CdS is another method to optimize the CBO of CdS/CZTSSe interface. Kato et al. from Solar Frontier K.K. indicated that the Zn and Se would diffuse out from CZTSe to CdS during deposition of CdS,therefore, the band edge of CdS would bend upward at the interface and the CBO at the CdS/CZTSe interface would be increased to 0.6 eV, which leads to S-shaped *J-V* curves with poor FF and efficiency.[24] They modified the CBD process of preparing CdS film to suppress the outdiffusion of Zn and Se, and thus the CBO was reduced to 0.3 eV, which is in the optimal range of 0-0.4 eV.[24] As a result, a CZTSSe submodule were prepared with a record efficiency of 10.8% and a low $V_{OC}$ deficit 561 mV.[24] Neuschitzer et al. optimized the CBD process of CdS by replacing of cadmium sulfate ($CdSO_4 \cdot H_2O$) with cadmium nitrate ($Cd(NO_3)_2$) as Cd precursor source to reduce the deep level acceptor-like traps states in the CdS and then reduce the CBO at CdS/CZTSe interface. As a result, the crossover and red kink of *J-V* curves were eliminated and the efficiency were improved from 7.0% to 8.2%.[25]

## 2.2. Zn(O,S)/CZTS(e) interfaces

CBO of CdS/CZTS(e) is sensitive to S/(S+Se) ratio. To optimize the band alignment between buffer layer and absorber layer, an alternative buffer layer, Zn(O,S), (Zn,Cd)S, $Zn_{1-x}Sn_xO_y$, $Zn_{1-x}Mg_xO$, and $In_2S_3$,etc., is used to optimize the band alignment at buffer/absorber interface. Among them, Zn(O,S) is studied as an alternative buffer layer due to the advantages of large and tunable band gap (2.6-3.8 eV), high transparency, low toxicity, and earth abundant raw materials.[26-28] Zn(O,S) buffer layer has been successfully used in CIGS solar cells, and world record efficiency 22.8% of CIGSSe solar cell with CBD-Zn(O,S) buffer layer has been achieved by Solar Frontier K.K., which is comparable with the devices with CdS-references.[29]

It is indicated that the CBO between ZnS and CZTSSe is spike like, which is as high as 1.1 eV (Fig. 3(a)). This high CBO will introduce a high barrier to electron flow and thus the photo-current is nearly zero.[30] However, the CBO between ZnO and CZTSSe is cliff like (Fig. 3(b)), thus the photo-current is much higher than that with ZnS buffer layer, and the $V_{OC}$ and $FF$ is much lower than that with CdS buffer layer. The band structure and band gap of Zn(O,S) are determined by the O/(S+O) ratio, as shown in Fig. 4.[31-32] So, it is crucial to optimize the band alignment between Zn(O,S) and CZTSSe via modifying the O/(S+O) ratio. Neuschitzer et al. studied the influence of thiourea (TU) concentration during the CBD process on the characteristics of Zn(O,S) layer as well as the influence on the device performances of CZTSe solar cells.[33] As shown in Fig. 5, the $J$-$V$ curves show a kink-like shape and the distortion of the $J$-$V$ curves are less pronounced with the decrease of TU concentration. The distortion of $J$-$V$ curves can be attributed to the high conduction band spike, which acts as a barrier for photo-generated electrons.[33] The O/(S+O) ratio in the Zn(O,S) buffer layer increase with the decrease of sulfur concentration in the CBO solution, and thus the barrier height decrease and even invert into cliff like for the sample with low TU concentration,causing the decrease even elimination of distortion (Fig. 5). As shown in Fig. 6, the distortion can be reduced and the $V_{OC}$, $J_{SC}$, and $FF$ increase significantly after light soaking, which can be attributed to the decrease of spike barrier after light soaking because of the photo-doping of Zn(O,S).[33] Grenet et al. also reported that the $J_{SC}$ is nearly zero with Zn(O,S) buffer layer due to the high spike barrier and the conversion efficiency increased significantly to 5.8% after light soaking due to the decrease of spike barrier.[34]

The Zn(O,S)/CZTSSe interface can also be improved by modifying the surface of CZTSSe absorber other than modifying the O/(S+O)ratio of Zn(O,S). Sakai et al. indicated that the efficiency of CZTS solar cells with Zn(O,S) buffer layer decreases with the Zn/Snratio of CZTS absorber.[35] Then, they reduced the surface Zn/Sn ratio from 1.1 to 0.96 via Tin-based chemical treatment (TCT) before CBD of Zn(O,S) buffer layer. As a result, a higher efficiency of 5.85% was achieved for a $5 \times 5$ cm$^2$ CZTS submodule.[36] The results suggest that the surface properties of absorbers are important to the properties of buffer/absorber interface and thus determine the device performance.

On the other hand, Zhang group in Nankai University China indicated the co-exist of ZnO and $Zn(OH)_2$ secondary phases in the Zn(O,S) layer will cause band fluctuations due to the large difference of conduction band edge between Zn(O,S) and ZnO, which is detrimental to the device performance, as shown in Fig. 7(a).[28] They proposed a treatment process of etching by concentrated ammonium and followed by soft annealing to eliminate ZnO and $Zn(OH)_2$ secondary phases in the Zn(O,S) layer.[28] As a result, the efficiency of the Zn(O,S)/CZTSe solar cells was improved from 1.17% to 7.2%. According to their study, the CBO between etched & annealed Zn(O,S) and CZTSe is spike like, as high as 1.22-1.28 eV, which will block the transportation of photo-generated electrons in theory. After study the low temperature transportation, they indicated that there is a defects energy level close to the Fermi level, which acts as a shortcut for the transportation of photo-generated electrons across the Zn(O,S) layer, as shown in Fig. 7(b).[28]

## 2.3. $Zn_{1-x}Cd_xS$/CZTS(e) interface

As we have mentioned, the CBO at CdS/CZTS and ZnS/CZTS is about -0.3 eV and +1.0 eV, respectively. Thus, $Zn_{1-x}Cd_xS$, a mixture of ZnS and CdS, can be used as buffer layer and the CBO between $Zn_{1-x}Cd_xS$ and CZTS can be optimized by modifying the Cd/(Zn+Cd) ratio. Bao et al. calculated the band offsets at $Zn_{1-x}Cd_xS$/CZTS interface by first principles and found that the CBO at $Zn_{1-x}Cd_xS$/CZTS interface is spike like and smaller than 0.3 eV, when the Cd/(Zn+Cd) ratio is 0.625 – 0.75.[37] Sun et al. prepared $Zn_{1-x}Cd_xS$ as buffer layer for CZTS solar cells by successive ionic layer adsorption and reaction (SILAR) method.[38] Their study indicated that all the performance parameters (including $V_{OC}$, $J_{SC}$, and $FF$) of CZTS solar cells with $Zn_{0.35}Cd_{0.65}S$ buffer are improved compared to CdS-references and the highest efficiency of 9.2% is achieved with the $V_{OC}$ of 747.8 mV, which is significantly improved about 100 mV compared to CdS-references.[38] The improvement is attributed to the spike like CBO (0.37 eV) at buffer/absorber interface, which will reduce the interface recombination. Messaoud et al. also used $Zn_{1-x}Cd_xS$ buffer layer for CZTSe solar cells and the $Zn_{1-x}Cd_xS$ buffers were deposited by chemical bath deposition.[39] It is indicated that the $V_{OC}$ and $FF$ decrease with the increase of Zn content in $Zn_{1-x}Cd_xS$ buffer due to the presence of $Zn(OH)_2$ and ZnO, which bring into the decrease of CBO. Thus, a Cd partial electrolyte (Cd PE) treatment before the deposition of $Zn_{1-}$

$_x$Cd$_x$S were adopt to form a thin Cd(OH)$_2$ layer (2-3 nm) to prevent the adsorption of the negatively charged Zn(O,OH)$_x$ particles, resulting in the decrease of interface recombination.[39] As a consequence, all the performance parameters of CZTSe solar cells with Cd PE/Zn$_{0.02}$Cd$_{0.98}$S buffer are improved compared to CdS-references, and a highest efficiency of 8.3% is achieved.

**2.4. (Zn,M)O/CZTS(e) interface (M= Sn, Mg)**

Zn$_{1-x}$Sn$_x$O$_y$ (ZTO) is also a promising buffer material for CZTS(e) solar cells. The conduction band alignment between ZTO and CZTS(e) can be optimized via tuning the Sn/(Zn+Sn) ratio. Platzer-Björkman et al. prepared ZTO buffer on CZTS absorber by atomic layer deposition (ALD) and optimized the band alignment between ZTO and CZTS by varying the deposition temperature.[40-41] It is worth noting that the highest efficiency of CZTS solar cells with ZTO buffer is 9.0%, which is higher than that of CdS-references.[41] Tajima et al. also used ZTO as buffer layer for CZTS solar cells and the best efficiency of CZTS solar cells with ZTO buffer is 5.7%, which is lower than its CdS-references.[42] They attribute the deterioration to the large lattice mismatch between ZTO and CZTS, which introduces interface recombination. To reduce the interface recombination, a ZTO/CdS double buffer layer was introduced.[42] Before the deposition of ZTO layer, a thin CdS buffer layer (10 nm) was deposited by CBD, followed by annealing in N$_2$ atmosphere. As a result, a high $V_{OC}$ of 0.81 eV was achieved and the efficiency was improved.

Zn$_{1-x}$Mg$_x$O (ZMO) is an attractive material as buffer layer due to its big and tunable band gap (3.3-7.7 eV). Besides, the conduction band alignment between ZMO and CZTS(e) can be optimized via tuning the Mg/(Zn+Mg) ratio. Hironiwa et al. prepared ZMO layer as buffer layer for CZTSSe solar cells, where the ZMO were deposited by co-sputtering of ZnO and MgO.[43-45] The efficiency of CZTSSe solar cells with ZMO buffer layer is significantly lower than that of CdS-references, which is attributed to the sputtering damage on CZTSSe absorber. To eliminate the sputtering damage, a ZMO/CdS double buffer layer was introduced.[43-45] As a result, the efficiency of solar cells with double buffer layer is improved to almost the same level of CdS-references. Besides, the efficiency of CIGS solar cells with double buffer layer is higher than the CdS-references.

**2.5. In$_2$S$_3$/CZTS(e) interface**

In$_2$S$_3$ is also a potential material as buffer layer for CZTS(e) solar cells. In$_2$S$_3$ is an indirect semiconductor with a band gap of 2.1 eV, which is slightly smaller than that of CdS (2.4 eV).[30, 46] However, the transparency of In$_2$S$_3$ is higher than that of CdS due to the indirect band gap of In$_2$S$_3$, which is beneficial to the improvement of $J_{SC}$.[30] Barkhous et al. indicated that the CBO between In$_2$S$_3$ and CZTSSe (for a sulfur content S/(S+Se) ~ 0.4 eV) is 0.15 ± 0.1 eV, the spike like in the optimal range.[30] The CZTSSe solar cells with In$_2$S$_3$ buffer layer yield a highest efficiency of 7.6%, which is comparable with the CdS-references. Yan et al. also used In$_2$S$_3$ as buffer layer in CZTS solar cells and conclude that the CBO between In$_2$S$_3$ and CZTS is 0.41 eV ± 0.1 eV, which is spike like and slightly higher than the optimal value. Thus, they indicated that the high CBO will block the flow of photo-generated electrons.[47] As a result, the CZTS solar cells with In$_2$S$_3$ buffer have a better $V_{OC}$ and lower $J_{SC}$ and $FF$ than the CdS-references. Yu et al. prepared In$_2$S$_3$ by thermal evaporation at different temperatures (30~200 °C). They found that the values of CBO between In$_2$S$_3$ and CZTS are dependent on the deposition temperature of In$_2$S$_3$ and variable in the range of 0.01 ~ 0.41 eV.[48] Besides, co-evaporated and sprayed In$_2$S$_3$ is used as buffer layer in CZTSe solar cells, and the efficiency of solar cells is around 5.7%.[49-50]

On the other hand, In$_2$S$_3$ can combined with CdS as double buffer layer to further improve the efficiency of CZTS(e) solar cells. Kim et al. indicated that a key reason for the low $V_{OC}$ of CZTSSe solar cells is the low majority carrier density and mobility product (i.e. the low conductivity) of the CZTSSe absorber.[51] The mobility of the CZTSSe absorbers is much lower than that of CIGS, thus the crucial to improve $V_{OC}$ is to increase the carrier concentration of absorber. Kim et al. prepared a thin In$_2$S$_3$ layer on CdS as a double buffer followed by rapid thermal annealing (RTA).[51] After RTA, In diffused into CdS and CZTSSe, leading to the formation of n-type doping defects In$_{Cd}$ in CdS and p-tye doping defects In$_{Sn}$ in CZTSSe, and resulting in the enhancement of carrier concentration of both CdS buffer and CZTSSe absorber. As a consequence, a record efficiency of 12.7% was achieved by employing an In$_2$S$_3$/CdS double buffer layer in CZTSSesolar cell.[51] Hiroi et al. also prepared a hybrid buffer layer with In$_2$S$_3$ deposited on CdS and an efficiency of 9.2% was obtained on CZTS submodule.[52] The improvement of device performance is attributed to the Cu and O diffusion as well as Cd and In diffusion, and the reduced carrier recombination at i-ZnO/buffer interface for the lower VBM at the interface.[52] Differently from Kim et al. and Hiroi et al.,[36, 51] Yan et al. prepared CdS/ In$_2$S$_3$ hybrid buffer with In$_2$S$_3$ inserted between CdS and CZTS, followed by a RTP annealing, yielding an improvement of $V_{OC}$ and efficiency from 5.5% to 6.6%.[53] It is indicated that the In can effectively diffuse into CdS and CZTS to increase the carrier concentrations of CdS and CZTS, and the carrier concentration of CZTS increased by on order of magnitude than CdS-references, resulting in the improvement of $V_{OC}$.[53]

In summary, the band alignment at CdS/CZTS(e) interface is not optimal and it is necessary to optimize the band alignment at buffer/CZTS(e) interface. First, the CBO can be optimized by modify the composition of CZTS(e) absorber, especially the surface composition. Secondly, the CBD process of CdS should be optimized to tailor the interfacial diffusion and reduce the interfacial defects. On the other hand, alternative buffer layers, including Zn(O,S), (Zn,Cd)S, $Zn_{1-x}Sn_xO_y$, $Zn_{1-x}Mg_xO$, and $In_2S_3$, etc., can be used to optimize the band alignment at buffer/absorber interface. However, the photovoltaic performances of CZTS(e) solar cells with an alternative buffer are usually not ideal and even worse than the CdS-references. To further improve the device performances of CZTS(e) solar cells with an alternative buffer, much more investigations are needed, including the optimization of the compositions and preparation processes of alternative buffer layers.

## 3. Mo/CZTS(e) interface

Mo films are the most commonly used back contact in CIGS solar cells because of a series advantage of Mo films as an excellent back contact. These merits include low resistivity, stability during growth at high temperature, and good adhesion between CIGS absorber and glass substrate.[54-58] Fig. 8 shows the schematic diagram of the band structure of the CIGS solar cells. As shown in Fig.8(a), the CIGS/Mo interface is a Schottky contact which would increase the series resistance $R_S$ and lead to the roll-over phenomenon in the I-V curve of the solar cell.[59] Generally, an interfacial layer of $MoSe_2$ can be formed between CIGS absorber and Mo back contact during the growth of CIGS film. As shown in Fig.8(b), the CIGS/$MoSe_2$ interface is Ohmic contact, whereas the $MoSe_2$/Mo interface is still a Schottky contact because of the large downward band bending at the $MoSe_2$/Mo interface. The $MoSe_2$ interfacial layer formed during the growth of CIGS films is usually very thin (10-100nm). Tunneling would occur across the $MoSe_2$/Mo interface when the thickness of $MoSe_2$ is very thin. Thus, the very thin $MoSe_2$ interfacial layer can act as a buffer layer to convert the Schottky contact to a quasi-Ohmic contact.[59-61] However, the transportation of holes will be suppressed when the $MoSe_2$ interfacial layer are very thick. Zhang's group indicated that that the back contact barrier height in CZTSe solar cells is 135 meV as MoSe$_2$ is very thick, while it is only 24 meV without MoSe$_2$ between Mo and CZTSe film.[62] Therefore, the thickness of MoSe$_2$ layer should be well-controlled to turn the back contact to a quasi-ohmic contact but not Schottky contact.

The structure of CZTS(e) solar cells is inherited from CIGS solar cells. Thus, Mo films are also widely used as back contact in CZTS(e) solar cells. However, different from CIGS films, the CZTS(e) films are easily to decompose at high temperature as shown below[63-64]

$$Cu_2ZnSnS(e)_4 \rightleftharpoons Cu_2S(e) + ZnS(e) + SnS(e) + \frac{1}{2}S(e)_2 \tag{1}$$

Besides, another decomposition process is easily to occur at the interface between CZTS(e) and Mo back contact at high temperature according to the following reaction[65]

$$2Cu_2ZnSnS(e)_4 + Mo \rightleftharpoons 2Cu_2S(e) + 2ZnS(e) + 2SnS(e) + MoS(e)_2 \tag{2}$$

It is calculated that change of the Gibbs energy of Reaction (2) is -150 kJ for CZTS and -100 kJ for CZTSe at 550 °C, which means that it is thermodynamically favorable for the decomposition at CZTS(e)/Mo interface and the MoS(e)$_2$ interfacial layer is easily to be formed between CZTS(e) absorber and Mo back contact.[63, 65] To suppress the formation of MoSe$_2$, Shin et al. modified the selenizing temperature.[66] They indicated that the thickness of MoSe$_2$ are reduced with the selenizing temperature decrease, whereas, the average grains size of CZTSe film are also reduced.[66] Another path to suppress the formation of MoSe$_2$ is selenizing precursors under low Se partial pressure. Yao et al. fabricated 8.2% CZTSe solar cells with a very thin MoSe$_2$ layer by selenizing electrodeposited Cu/Sn/Zn precursor under low Se pressure.[67] However, deep defects (e.g. Se deficiency) are formed when the Se partial pressure is low during selenizing at high temperature.[66]

To promote the growth of grains, suppress the decomposition according to Reaction (1), and avoid the formation of deep defects (e.g. Se deficiency), CZTS(e) absorbers are usually prepared at high temperature under high S(e) and SnS(e) pressure.[65-66, 68-70] As a result, an excessively thick MoS(e)$_2$ interfacial layer, which would increase the contact resistance between CZTS(e) absorber and Mo back contact, is easily to be formed according to Reaction (2).[66, 69] Thus, one of the challenges for the improvement of the current conversion efficiencies is to reduce the thickness of MoSe$_2$ interfacial layer. After studying the kinetics of MoSe$_2$ interfacial layer formation during annealing CZTSe thin films under Se ambient, Shine et al. indicated that the formation of MoSe$_2$ is modeled with three step processes, as shown in Fig.9—diffusion of Se through CZTSe film, diffusion of Se through the already formed MoSe$_2$, and then reaction between Se and Mo.[69] It reveals that the key to suppress the formation of MoSe$_2$ is to block the diffusion of Se. Thus, there are following several ways to suppress the formation of MoSe$_2$.

In accordance with the characteristic that Se vapor firstly diffuse through CZTS(e) film, an intermediate layer, such as TiN, TiB$_2$, MoO$_2$, and ZnO etc., inserted between CZTS(e)absorber and Mo back contact are introduced as a S(e) diffusion barrier to suppress the formation of MoS(e)$_2$.[64, 66, 71-72]

TiN is used as the barrier material to prevent the S(e) diffusion for its stability and minor diffusion at high temperature. Shin et al. reduced the thickness of MoSe$_2$ layer from 1300 nm to 220 nm by a thin TiN diffusion barrier (~20 nm) and thereby the series resistance decreased from 3.4 • cm$^2$ to 1.8 • cm$^2$ and the efficiency increased from 2.95% to 8.9%.[66] However, Scragg et al. found that TiN intermediate layer would induce a rather high series resistance, causing the decrease of fill factor.[64] Therefore, contrast to the results of Shin,[66] the efficiency of CZTS solar cells was reduced when a TiN intermediate layer was introduced although the MoS$_2$ were out of growth.[64] Schnabel et al. further investigated the influence of different back contact configurations of Mo and TiN (shown as Fig. 10) on the formation of MoS$_x$Se$_{2-x}$, the crystallinity of the absorber and thereby the performance of solar cell devices.[72] They found that the thin TiN intermediate layer can effectively inhibit the formation of MoS$_x$Se$_{2-x}$, but is detrimental to the crystallinity of the absorber layer. The photovoltaic performances of solar cell demonstrated that the existence of MoS$_x$Se$_{2-x}$ is necessary for achieving high $V_{OC}$. Thus, the best performance solar cells were the ones using a combined Mo/TiN/Mo stack as back contact, which can easily control the thickness of MoS$_x$Se$_{2-x}$ layer by the thickness of the upper Mo layer.

In addition to TiN, TiB$_2$ is introduced as the intermediate layer. Liu et al. inhibited the formation of MoS$_2$ layer with TiB$_2$ film as intermediate layer in CZTS thin film solar cells.[71] As a result, the series resistance was greatly reduced and thereby the efficiency of solar cells were increased from 3.06% to 4.40% with the improvement of $J_{SC}$ and FF. However, the TiB$_2$ intermediate layer would degrade the absorber's crystallinity, which would be detrimental to the device performance especially $V_{OC}$.[71]

Besides, Duchatelet et al. prepared a MoO$_2$ layer by oxidizing the Mo coated glass substrates to suppress the formation of MoSe$_2$.[73] It is indicated that the Gibbs free energy of the reaction of MoO$_2$ with selenium is +294 KJ/mol which is thermodynamically unfavorable.[73] As a result, the MoO$_2$ can passivate Mo against selenization, causing the significant decrease of thickness of

MoSe$_2$ layer.[73] It is known that MoO$_2$ exhibit a metallic electrical conductivity,[74-75] and thus the presence of MoO$_2$ with a thin MoSe$_2$ has little effect on the sheet resistance of back contact.[73] Lopez-Marino et al. found that the presence of MoO$_2$ will change the preferred orientation of MoSe$_2$ from (100) peak to (002) peak, i.e. the Se-Mo-Se sheets changing from perpendicular to parallel to Mo substrate, as shown in Fig.11(a) and (c).[76] The MoSe$_2$ with Se-Mo-Se parallel to Mo substrate can act as a natural Se barrier and thus the thickness of MoSe$_2$ are reduced.[76] However, the conductivity of MoSe$_2$ is anisotropic, and the electrical conductivity along the Se-Mo-Se sheets is two orders of magnitude bigger than that along the direction perpendicular to the Se-Mo-Se sheets, i.e. the electrical conductivity of MoSe$_2$ with Se-Mo-Se sheets parallel to Mo substrate is very poor.[77] To prepare sufficient MoSe$_2$ promote an Ohmic contact between Mo and absorber, another top sacrificial Mo cap layer (20-70 nm) was deposited on the MoO$_2$ layer.[76] The thickness of MoSe$_2$ can be tailored by modifying the thickness of the top sacrificial Mo cap layer and the photovoltaic performances can be improved. Besides, the MoO$_2$ layer can not only reduce the thickness of MoSe$_2$ effectively but also can obviously promote the grain growth, and thus the shunt resistance ($R_{Sh}$) was increased.[76]

An intermediate ZnO layer on Mo back contact can also prevent the interaction between CZTS(e) absorber and Mo film during the high temperature annealing process and thereby the decomposition of CZTS(e) would be inhibited. In 2013, López-Marinoet al. introduced an ultrathin (10nm) ZnO intermediate layer to minimize/eliminate the decomposition reaction between CZTSe and Mo interface.[78] Consequently, the Cu$_x$Se binary species were eliminated and the density of voids was reduced. As a result, the series resistance were significantly decreased and thereby the efficiency increased from 2.5% to 6.0%.[78] Li et al. and Liu et al. also inhibited the formation of MoS$_2$ by introducing an ultrathin (10 nm) ZnO intermediate layer.[79-80]

It needs to be indicated that the path to suppress the formation of MoS(e)$_2$ by inserting an intermediate layer can work well in some extent. However, it will make the process complexity and the series resistance of solar cells cannot be decreased to a low level. To simplify the process to suppress the formation of MoSe$_2$ and decrease the series resistance of solar cells, Zhang's group presented two means to suppress the formation of MoSe$_2$ successfully without any additional intermediate layer used.[62, 81] One is that the metal precursors were annealed at 300 $^{\circ}$C under Ar atmosphere to form a dense temporary alloy layer, the diffusion of Se was blocked in the following selenization process at high temperature and the thickness of $MoSe_2$ interfacial layer were reduced from 1200 nm to less than 10 nm. The series resistance was reduced from 3.4 $\Omega \cdot cm^2$ to 0.61 $\Omega \cdot cm^2$, and the conversion efficiency were improved from 5.6% to 8.7%.[62] The other is that a thin (103) peak preferred $Mo(S,Se)_2$ layer with tilted Se-Mo-Se sheets were prepared by modifying the surface morphology of Mo films.[81] As shown in Fig. 11 (d), the tilted Se-Mo-Se sheets can act as a natural barrier to inhibit the formation of $Mo(S,Se)_2$ interfacial layer and provide a good electrical conductivity along the Se-Mo-Se sheets.[81] As a result, the thickness of $Mo(S,Se)_2$ sharply decreased from 1500 nm to 200 nm with the surface morphology change of Mo back contact, which resulting in the decrease of series resistance of $Cu_2ZnSn(S,Se)_4$ solar cells from 2.94 $\Omega \cdot cm^2$ to 0.49 $\Omega \cdot cm^2$, and the increase of conversion efficiency of $Cu_2ZnSn(S,Se)_4$ solar cells from 6.98% to 9.04%.

In summary, a thin $MoS(e)_2$ layer between CZTS(e) absorber and Mo back contact can act as a buffer layer to convert the Schottky contact to a quasi-ohmic contact. However, an over thick $MoS(e)_2$ layer is easily to be formed during selenaziton process at high temperature under high S(e) pressure, which will suppress the transportation of holes. Thus, it is crucial to take effective measures to tailor the thickness of $MoS(e)_2$. The formation of $MoS(e)_2$ can be suppressed by a dense temporary precursor, a thin (103) peak preferred $Mo(S,Se)_2$ layer with tilted Se-Mo-Se sheets, or an intermediate barrier layer, including TiN, $TiB_2$, and $MoO_2$, etc..

## 4. Passivation

### 4.1. Passivation of rear interface

To reduce the rear recombination at the semiconductor-metal interface to increase the $V_{OC}$, and meanwhile to improve the $J_{SC}$ by increasing the rear reflection, a dielectric layer with patterned local contacts has been introduced between CZTS(e) and Mo to passivate the rear interface. The dielectric layer, such as $SiO_2$[82-84], $SiC_X$[85], $SiN_X$[83], $Al_2O_3$[84], or $Al_2O_3(SiO)/SiN_X$[86] have been used as a passivation layer in Si solar cells for a long time. Recently, the dielectric layer has been adopted in CZTS(e) solar cells. Vermang et al. used an $Al_2O_3$ with nanosized point openings to passivate ultrathin CZTS solar cells.[87] The results indicated that the charge carrier recombination and the secondary phases (SnS and ZnS) were reduced. As a consequence, compared with CZTS solar cells without rear passivation at CZTS/Mo interface, photovoltaic performances of the CZTS solar cells with an $Al_2O_3$ rear passivation layer were obviously increased.[87] Similarly, Kim et al. prepared a dielectric layer

($Al_2O_3$ or $SiO_2$) with nano-patterned local contacts by nanosphere lithography to passivate the Mo/CZTSe interface, as shown in Fig. 12.[88] All photovoltaic parameters were significantly improved and the luminescence intensity of LT-PL spectra was increased for the CZTSe solar cells with a dielectric layer, which indicates that the $Al_2O_3$ or $SiO_2$ dielectric layer passivated the bottom interface effectively.[88] Different from the study of Vermang[87] and Kim[88], Liu et al. sputtered a $Al_2O_3$ intermediate layer on the Mo back contact, and then the $Al_2O_3$ intermediate layer turns into a self-organized nanopattern with a nanoscale opening between CZTS and $MoS_2$ for their connection.[89] The $Al_2O_3$ intermediate layer prevents the interface reaction between CZTS and Mo at the initial stage of sulfurization, and thus the thickness of $MoS_2$ is reduced, the voids and ZnS secondary phase at the back contact region are eliminated. And then the initially continuous $Al_2O_3$ intermediate layer breaks into discontinuous aggregates, producing self-organized nanopattern with nanoscale opening.[89] The CZTS absorber and $MoS_2$ can connect directly at the openings and it is beneficial to the electrical contact and avoids high series resistance of the solar cell devices. Besides, the crystallinity of CZTS absorber becomes better, and the minority lifetime is increased. As a result, the conversion efficiency of the ultrathin CZTS solar cells is improved from 7.34% to 8.65%.[89]

An electrostatic field at the back contact can also prevent rear recombination, which can be created by depositing a high work function (HWF) material, such as $MoO_3$, $WO_3$, etc., between CZTSSe absorber and Mo back contact. It is indicated that the work function of $MoO_3$ is 6.5 eV, which is much higher than that of CZTSSe (5.2 eV).[90] Thus, electrons flow from the lower work function CZTSSe to the higher work function $MoO_3$ to equilibrate the Fermi level while $MoO_3$ and CZTSSe are contacted and thus an electrostatic field is created. Thereby, electrons will be driven to the front hetero-junction, holes will be attracted to the back contact, and the electron-hole recombination at back contact will be reduced, as shown in Fig. 13. Antunez et al. presented a method to modify the back contact by exfoliating the fully CZTSSe devices from Mo/glass substrate, followed by the deposition of HWF $MoO_3$ and reflective Au contact.[90] The wxAMPS simulation results indicate that the $V_{OC}$ increases with the absorber thickness decrease while the high work function material ($MoO_3$) are applied and it is true for actual devices.[90] It is because that more electron-hole pairs are created at and near the back contact with the decrease of the thickness of absorber. The carrier recombination is reduced when an electrostatic field created by HWF $MoO_3$. It is indicated that the $V_{OC}$ increases of up to 49 mV for the non-etched samples, and even up to 61 mV for the bromin-methanol etched samples with 1 μm absorber thickness.[90] Ranjbar et al. also introduced an ultra-thin $MoO_3$ layer between CZTSe absorber and Mo back contact and found that the minority carrier life time and $V_{OC}$ are increased because of the decrease of rear recombination.[91]

Besides, to minimize/eliminate the blocking of carrier transportation by voids, Zhou et al. introduced an ultrathin carbon layer between CZTS and Mo layer.[92] It is indicated that the carbon sticks on the inner wall of voids and connects the CZTS and $MoS_2$ at the void region after sulphuration, which will boost the transportation of free carriers and improve the $J_{SC}$.[92] Besides, the introduction of the ultrathin carbon intermediate layer has no obvious influence on the thickness of $MoSe_2$ and the crystallinity of CZTS film, and thereby the $V_{OC}$ and FF are nearly unchanged.[92]

### 4.2. Passivation of front interface

An ultrathin dielectric film, such as $TiO_2$ and $Al_2O_3$, can be introduced between CdS and CZTSSe, acting as a passivation layer to reduce the recombination at CdS/CZTSSe interface.[88, 93-96] Wu et al. introduced an atomic layer deposition (ALD) ultrathin $TiO_2$ layer between CdS and CZTSSe to reduce the interface recombination.[93] It is indicated that the activation energy ($E_A$) of the main recombination process of the CZTSSe device with $TiO_2$ passivation layer is ca. 40 meV higher than that without a passivation layer, which suggests that the interfacial recombination is reduced for the passivation of $TiO_2$ layer.[93] Besides, the DLCP and CV profiling suggests the density of defects at CdS/CZTSSe interface is reduced for the CZTSSe devices with $TiO_2$ passivation layer.[93] As a result, the $V_{OC}$, FF, and efficiency of CZTSSe solar cells with $TiO_2$ layer are improved. However, the $J_{SC}$ is decreased slightly due to the impact of insulating $TiO_2$ layer that the charge transportation is suppressed.[93] Therefore, it is critical to optimize the thickness of the passivation layer to minimize the blocking of charge transportation. It is convenient to control the thickness of the ultrathin passivation layer by ALD, while it is difficult for other deposition technique. Ranjbar et al. adopted a solution processed $TiO_2$ layer to passivate the front interface.[94] The introduction of $TiO_2$ layer improved the $V_{OC}$, whereas, the FF and efficiency decreased which is because of a high barrier at the conduction band by the introduction of $TiO_2$ layer and therefore the charge transportation is suppressed.[94] To boost the charge transportation, opens were created by an additional KCN treatment after $TiO_2$ layer deposited, and the $J_{SC}$ and $FF$ are also improved, and the efficiency was improved from 5.4% to 6.9%.[94]

In addition, introduction of $Al_2O_3$ layer is an effective interface-passivation strategy to improve device performance of CZTS(e) solar cells. Lee et al. introduced an ultrathin ALD-$Al_2O_3$ film as a passivation layer.[95] They indicated that $Al_2O_3$ also penetrated into grain boundaries and pinholes of CZTSSe film to passivate these regions.[95] As the i-ZnO layer were replaced by $Al_2O_3$, the sputtering damage to CdS surface during TCO preparation was reduced and consequently the $V_{OC}$ were improved ca. 5%.[95] By introducing the aforementioned type of $Al_2O_3$ layer, the CZTSSe device performances were improved significantly and the highest conversion efficiency is 11.5%.[95] It is noticeable that the $Al_2O_3$ film can be etched by $NH_4OH$ during CdS deposition, which is different from the stable of $TiO_2$ film in the CBD environment.[94] Therefore, the series resistance ($R_S$) is almost the same with the thickness increase of $Al_2O_3$ layer between CZTS and CdS, whereas the $R_S$ obviously increases with the thickness increase of $Al_2O_3$ layer between CdS and ITO.[95] Erkan et al. also introduced ALD-$Al_2O_3$ layer between CdS and CZTSSe and found that the density of acceptor-like defects is reduced and the depletion width is widened. As a result, $V_{OC}$ are improved.[96]

Kim et al. indicated that there is a strong accumulation of Cu and some deficiency of Cd near the interface of CdS/CZTSe, i.e. the intermixing of Cu and Cd is severe between CdS and CZTSe during the CdS deposition (shown in Fig. 14(a)).[88] In CIGS solar cells, the Cd doping and the formation of $Cd_{Cu}$ donor defects near the CdS/CIGS interface will cause a large band bending and convert the CIGS surface to n type which is beneficial to the separation of photo-generated carries.[97-98] In kesterite solar cells, however, it is disclosed that neutral defect cluster ($Cd_{Cu}$ + $V_{Cu}$) is most easily formed during CdS deposition on CZTS(e) film, and a small amount of donor defect $Cd_{Cu}$ and neutral defect $Cd_{Zn}$ are formed.[99] Thus, the surface of CZTS(e) film is difficult to convert to n-type, and thereby the benefits of intermixing of Cu and Cd near CdS/CZTS(e) interface is ambiguous. Kim et al. introduced a 5 nm $Al_2O_3$ layer between CdS and CZTSe, which effectively blocked the inter-diffusion of Cu and Cd (shown in Fig. 14(b)).[88] Besides, the intensity increase of PL spectra indicated that the non-radiative recombination is suppressed for the sample with $Al_2O_3$ passivation layer (shown in Fig. 14(d)). Finally, $V_{OC}$ and $FF$ were significantly improved, which indicates that the server intermixing between CdS and CZTSe is detrimental to photovoltaic performances of CZTSe solar cells.[88]

The density of interface defects can be reduced by epitaxial growth of buffer layer on CZTS(e) absorber. Liu et al. achieved the epitaxial growth of CdS on CZTS via sulfurization in combined SnS & S atmosphere, or sulfurization in only S atmosphere with surface defects but passivated by air annealing.[100] Their results indicated that the epitaxial growth of CdS leads to a low interfacial defects density and thereby low recombination rate. So, the minority lifetime is increased and the device efficiency of CZTS solar cells are improved to 8.76%.[100]

Although the density of interface defects can be reduced by epitaxial growth of CdS on CZTS, the lattice mismatch between CdS and CZTS is much larger (ca. 7%) than that between CdS and CIGS (ca. 1.5%) or CZTSe (ca. 2.4%), which is detrimental to the epitaxial growth CdS on CZTS.[101] $CeO_2$ has a nearly perfect lattice match with CZTS (ca. 0.4%) and the interface defects can be decreased significantly. Besides, the conduction band offset (CBO) between $CeO_2$ and CZTS is -0.12 ± 0.20 eV, which is only slightly below the optimal CBO range, and it is more favorable than the CdS/CZTS CBO. Thus, Crovetto et al. proposed that the interface recombination can be reduced by the introduction of a $CeO_2$ layer between CdS and CZTS.[101] As a result, the $V_{OC}$ is improved. However, the electron effective mass of $CeO_2$ is very high, and thereby the electron mobility is very low.[101] As a result, the photocurrent transportation is suppressed and the $J_{SC}$ is reduced. Thus, the thickness of $CeO_2$ should be ultrathin and controlled precisely. It is better that the ultrathin $CeO_2$ layer is prepared via atomic layer deposition.

In summary, the interfacial defects will increase the interfacial recombination, so the passivation of interface is a key point for high efficient CZTS(e) solar cells. The interface can be passivated by an ultrathin dielectric layer, including $Al_2O_3$, $TiO_2$, and $SiO_2$ etc., and the epitaxial growth of buffer on CZTS(e) absorber in some extent. AHWF material, including $MoO_3$ and $WO_3$, etc., can be inserted between CZTSSe absorber and Mo back contact to form an electrostatic field at back contact, and thereby promote the transportation of charge electrons. It needs to further explore the passivation for the interface to improve the transportation of charge carriers.

## 5. Etching of secondary phases

It is calculated that the phase stable regions of CZTS and CZTSe are much smaller than that of $CuInSe_2$ (CISe) because of the much more competing secondary phases, as shown in Fig. 15.[102] Compared with CISe, although the stable chemical potential range of $\mu_{Cu}$ (-0.4 to 0 eV) in kesteriteis slightly smaller than that of CISe (-0.5 to 0), the range of $\mu_{Zn}$ (ca. 0.2 eV) and $\mu_{Sn}$ (ca. 0.6 eV) are much smaller than that of $\mu_{In}$ (ca. 1.0 eV).[102] As a result, various secondary phases, such as CuS(e), Cu$_2$S(e), Cu$_2$SnS(e)$_3$, ZnS(e), SnS(e), and SnS(e)$_2$, are easily to coexist in CZTS(e) absorber. A deficit of Zn would lead to the segregationof Cu$_2$SnS(e)$_3$, SnS(e), and SnS(e)$_2$.[102] The band gap of Cu$_2$SnS(e)$_3$ and SnS(e) are smaller than that of CZTS(e), which will increase the $V_{OC}$ deficit. The conduction band edge energy of SnS(e)$_2$ is much lower than that of CZTS(e).[103-104] Therefore, the SnS(e)$_2$ secondary phases tend to trap minority charge carriers, produce shunt paths, and thereby deteriorate the $V_{OC}$ and *FF* of CZTS(e) devices.[105-107] On the other hand, an excess of Zn will lead to the segregation of ZnS(e). ZnS(e) secondary phases are less detrimental to device performance as they are on the back of CZTS(e). However, the presence of ZnS(e) on the surface of CZTS(e) will inhibit the transportation of charge carriers because of the big spike CBO (1.32 eV) between ZnS(e) and CZTS(e), causing the decrease of $J_{SC}$ and the increase of $R_S$.[108-112] Besides, the presence of Cu$_x$S(e) secondary phases can introduce shunt paths for its high conductivity.[107, 113-115] It is indicated that the secondary phases tend to accumulate near the top or bottom of CZTS(e) absorber, which will increase the interface recombination and thereby deteriorate the performance of CZTS(e) solar cells.[116-117] Therefore, the composition control, the avoidance of the formation of secondary phases, and the removal of secondary phases are crucial for high efficient CZTS(e) solar cells. The secondary phases at the bottom of CZTS(e) can be eliminated via suppressing the reaction between CZTS(e) and Mo by an intermediate layer, such as TiN, Al$_2$O$_3$, and ZnO, etc., or modifying the sulfuration/selenization process.[64, 78-80, 87, 118] As for the secondary phases on the front surface of kesterite film, chemical etching is an effective way to remove them.

Cu$_x$S(e) secondary phases are usually removed by potassium cyanide (KCN) etching.[119-122] The measurement of Kelvin probe force microscopy (KPFM) and Raman scattering spectroscopy suggested that the Cu$_x$S(e) can be removed completely by KCN etching.[119-121] KCN is toxic and hazardous, thus the usage content should be very small. Buffière adopted a 5 % KCN aqueous solution containing 0.5 wt % KOH for safety reasons.[122] The *FF* of CZTSe solar cell are improved after 30 s of etching. However, they found that elements Zn, Sn, and Se are preferentially removed from CZTSe surface by KOH, causing an increase of Cu content at CZTSe surface after long etching time.[122] The Cu-rich surface will increase the recombination rate at CdS/CZTSe interface and thereby deteriorate the *FF* of CZTSe solar cells. Therefore, the

KCN etching time should be less than 30 s and/ or the concentration of KOH should be reduced to improve the performance of CZTSe solar cells.[122] Other than the removal of $Cu_xS(e)$ phases, KCN etching can improve the performance of CZTS(e) solar cells in other ways. Bär et al. indicated that the surface composition will be changed after KCN etching. Therefore, the surface band gap will be increased from 1.53 ± 0.15 eV to 1.91 ± 0.15 eV, which will enhance the cliff-like conduction band offset at CdS/CZTS.[16] Then the energetic barrier for charge carrier recombination will be increased and thereby the $V_{OC}$ will be improved.[16] Besides, KCN etching can also remove the surface overlayer (including native oxides and airborne contamination) completely, which will reduce the density of contamination-induced surface defects and the interface recombination.[123] In addition, Pinto et al. reported the $Cu_{2-x}(S,Se)$ secondary phase can be removed completely by a safer chemical mixture of ethylenediamine and 2-mercaptoethanol.[124] It is indicated that $H_2O_2$ can also effectively remove $Cu_xS$ phases following by[125]

$$Cu^{2+}S^{2-} + 4H_2O_2 \rightarrow CuSO_4 + 4H_2O \qquad (3)$$

$$Cu_2^+S^{2-} + 4H_2O_2 \rightarrow Cu_2SO_4 + 4H_2O \qquad (4)$$

ZnS secondary phases can be selectively removed by HCl solution.[126] Fairbrother et al. reported the efficiency of CZTS solar cells are improved from 2.7% to 5.2% after HCl etching.[126] As for ZnSe secondary phase, Mousel et al. adopted the HCl solution to etch the surface ZnSe secondary phases and improved the device efficiency from about 4% to above 5%.[127] However, ZnSe secondary phases are mostly removed via an oxidizing route using $KMnO_4/H_2SO_4$ followed by a $Na_2S$ etching.[128] Firstly, ZnSe phases are oxided to $Se^0$, and then the $Se^0$ phases are removed by $Na_2S$, as shown in Table 1. López-Marino used three different oxidizing agents, $H_2O_2$, $KMnO_4$, and $K_2Cr_2O_7$, to etch ZnSe and found that $KMnO_4$ is more effective than others.[128] The device performances are imprvoed significantly after ZnSe etching. The improvement of $J_{SC}$ and series resistance can be attributed to minimizing the blocking of charge carriers transportationdue to the ZnSe secondary phases. Besides, the improvement of $V_{OC}$, $FF$, and $R_{sh}$ can be ascribed to a chemical passivation of the CZTSe absorber surface.

SnS(e) and SnS(e)$_2$ can be removed from the CZTS(e) surface by $(NH_4)_2S$ etching.[106, 109, 129-131] The removal of Sn-Se phases were demonstrated by comparing the Sn content, Sn-Se XRD peaks, and Raman signals before and after etching.[129] Xie et al. indicated that the $(NH_4)_2S$ solution may also passivate the surface of CZTSe absorber via removal of surface oxides and formation of S passivated species other than removal of Sn-Se scondary phases. As a result, all *J-V* parameters of CZTSe solar cells are improved and the conversion efficiency increases between 20 and 65% after $(NH_4)_2S$ etching.[129] However, it is indicated that the longer etching time tend to deteriorate the device performance because of the preferred etching of Se, Sn, and Zn, causing the increase of Cu content at the CdS/CZTSe interface.[130] Wei et al. presented the removal of SnS secondary phases using $Na_2S$ solution, which were certified by Raman scattering mapping and PL spectra.[132] Ren et al. indicated that the SnS secondary phases on CZTS surface are detrimental to solar cells, whereas the SnS on CZTS rear can passivate the back contact, because of the spike like CBO between CZTS and SnS, limiting electron transportation toward the back contact, as shown in Fig. 16.[133] Therefore, the formation location of SnS secondary phase should be carefully controlled. As for $Cu_2SnS(e)_3$ secondary phases, they can be etched in Br in methanol ($Br_2$-MeOH).[127, 134]

In summary, the stable region of CZTS and CZTSe are too small, which causes the presence of secondary phases. The secondary phases are detrimental to the photovoltaic performances of CZTS(e) solar cells. Chemical etching is an effective way to remove the secondary phases on the surface of CZTS(e) absorbers. The photovoltaic performances of CZTS(e) solar cells can be improved after the removal of secondary phases.

## 6. Summary and perspective

The efficiency of CZTSSe solar cells has reached at 12.6%, which is fabricated by solution method. However, the efficiency cannot be improved for several years because of high Voc deficit, which is related to many aspects. Interface is one of such aspects. In this review, we discussed the problems at interfaces of CZTS(e) solar cells, which are detrimental to the photovoltaic performances, and the corresponding solutions to overcome the problems and improve the photovoltaic performances. First, the conduction band alignment at CdS/CZTS(e) interface is not optimal. The CBO between CdS and CZTS or CZTSe is cliff like or a slightly bigger spike barrier, respectively, which is detrimental to $V_{OC}$ or $J_{SC}$, respectively. The CBO can be optimized by modify the composition of CZTS(e) absorber or adopt an alternative buffer layer, including Zn(O,S), (Zn,Cd)S, $Zn_{1-x}Sn_xO_y$, $Zn_{1-x}Mg_xO$, and $In_2S_3$,etc.. Secondary, over thick $MoS(e)_2$ layer is easily to be formed during selenization/sulphuration process at high temperature under high Se/S pressure, so the thickness of $MoS(e)_2$ layer between CZTS(e) absorber and Mo back contact should be tailored. The formation of $MoS(e)_2$ can be suppressed by a dense temporary precursor, a MoS(e)$_2$ layer with titled Se-Mo-Se sheets, or an intermediate barrier layer, including TiN, TiB$_2$, and MoO$_2$, etc.. Thirdly, the interfacial defects will increase the interfacial recombination, so interfacial passivation is necessary. A dielectric layer, including Al$_2$O$_3$, TiO$_2$, and SiO$_2$ etc., can be used to passivate the front and rear interface. An electrostatic field at back contact, created by depositing a high work function (HWF) material (such as MoO$_3$, WO$_3$, etc.) between CZTS(e) absorber and Mo back contact, can also prevent rear recombination. Besides, the epitaxial growth of buffer on CZTS(e) absorber will reduce the interfacial recombination in some extent. Finally, secondary phases are easily to be segregated due to the small stable regions of CZTS(e). Chemical etching is an effective way to remove the secondary phases on the surface of CZTS(e) absorbers. The photovoltaic performances of CZTS(e) solar cells can be improved after the removal of secondary phases.

Although various methods have been adopted to improve the interfacial properties, the $V_{OC}$-deficit and efficiency of CZTS(e) solar cells still have a big gap with their S-Q limit and that of CIGS solar cells. So, more research is necessary, including adopting a new preparation process (like MOCVD) of CZTS(e) absorber to improve the interfacial properties, adopting a new buffer layer to optimize the band alignment, and adopting a new passivation method and passivation material to passivate the defects at interfaces, etc..The main advantage of MOCVD over other growth techniques is the low growth temperature, thereby the decomposition of CZTS(e) absorber can be avoided, resulting in the high purity of the material and the absence of secondary phases.

Figure captions

**Fig. 1.**(color online) Temperature dependence of $V_{OC}$ for CZTSSe and CIGSSe solar cells. Reproduced with permission from Ref. [10].

**Fig. 2.**(color online) Schema of band alignment at buffer/absorber interface: (a) cliff like, (b) and spike like. Spike and cliff are barriers for photo-generated and injection electrons, respectively. Reproduced with permission from Ref. [15].

**Fig. 3.**(color online) Schematic of the band alignment between CZTSSe and (a) ZnS and (b) ZnO, respectively. Reproduced with permission from Ref. [30].

**Fig. 4.**(color online) Band energy diagram of Zn(O,S) as a function of O/(S+O). Reproduced with permission from Ref. [31-32].

**Fig. 5.**(color online) Light and dark *J-V* curves of CZTSe solar cells with Zn(O,S) buffer layer prepared with different TU concentration during the CBD process and CdS as a reference buffer layer. Reproduced with permission from Ref. [33].

**Fig. 6.**(color online) Light *J-V* curves of CZTSe solar cells with Zn(O,S) buffer layer prepared with different TU concentration during the CBD process before and after light soaking for 260 min. Reproduced with permission from Ref. [33].

**Fig. 7.**(color online) The schematic diagrams of band alignments of the Zn(O,S)/CZTSe interface for (a) the Untreated sample and (b) the Etched&annealed sample. Reproduced with permission from Ref. [28].

**Fig. 8.**(color online) Schematic diagram of the band structure of the CIGS solar cells: (a) without $MoSe_2$ layer, and (b) with $MoSe_2$ layer.

**Fig. 9.**(color online) Schematic of excess Se concentration profile across CZTSe/$MoSe_2$/Mo during annealing in a steady-state. Reproduced with permission from Ref. [69].

**Fig. 10.**(color online) Schematic drawings of the different back contact configurations. Reproduced with permission from Ref. [72].

**Fig. 11.** (color online) Crystal structures of MoSe$_2$ along different direction, which is created with CrystalMaker software: (a) (100) preferred orientation with Se-Mo-Se sheets perpendicular to the substrate, (b) (110) preferred orientation with Se-Mo-Se sheets perpendicular to the substrate, (c) (002) preferred orientation with Se-Mo-Se sheets parallel to the substrate, and (d) (103) preferred orientation with Se-Mo-Se sheets tilted to the substrate. Reproduced with permission from Ref. [81].

**Fig. 12.** (color online) (a) Schematic drawings of passivation layer with nano-patterned local contacts. (b) SEM images of contact holes on Mo after nanosphere lithography. (c) Cross-sectional TEM image of a bottom Al$_2$O$_3$ passivation layer (white dotted square) with a local contact between CZTSe and MoSe$_2$. Insets are Fourier transform images taken by the area of near the surface (yellow dotted square) and the bulk (red dotted square), respectively. (d) *J-V* characteristics for the CZTSe solar cells with and wihout a (d) Al$_2$O$_3$ and (e) SiO$_2$ passivation layer at the CZTSe/Mo interface. (f) 10 K photoluminescence spectra of CZTSe solar cells with and without an Al$_2$O$_3$ passivation layer at the CZTSe/Mo interface. Reproduced with permission from Ref. [88].

**Fig. 13.** (color online) Schematic band structure schematic of a CZTSSe solar cell with a high work function material between CZTSSe absorber and Mo back contact.

**Fig. 14.** (color online) HAADF-STEM images and corresponding EDS mapping of Cu, Cd, and S of CZTSe solar cells (a) without and (b) with an Al$_2$O$_3$ passivation layer at the CZTSe/CdS interface. (c) Changes in *V*oc and efficiency as a function of Al$_2$O$_3$ thickness (solid lines are guides to eyes and a dashed line indicates a respective reference). (d) 10 K PL spectra of CZTSe solar cells with (purple) and without (green) a top Al$_2$O$_3$ passivation layer at the CZTSe/CdS interface. Reproduced with permission from Ref. [88].

**Fig. 15.** (color online) The calculated phase stable region of CISe (left), CZTS (center), and CZTSe (right). Reproduced with permission from Ref. [102].

**Fig. 16.** (color online) (a) Band edges at the interface of CZTS/SnS and (b) charge carrier transport paths close to CZTS/SnS interface. Reproduced with permission from Ref. [133].

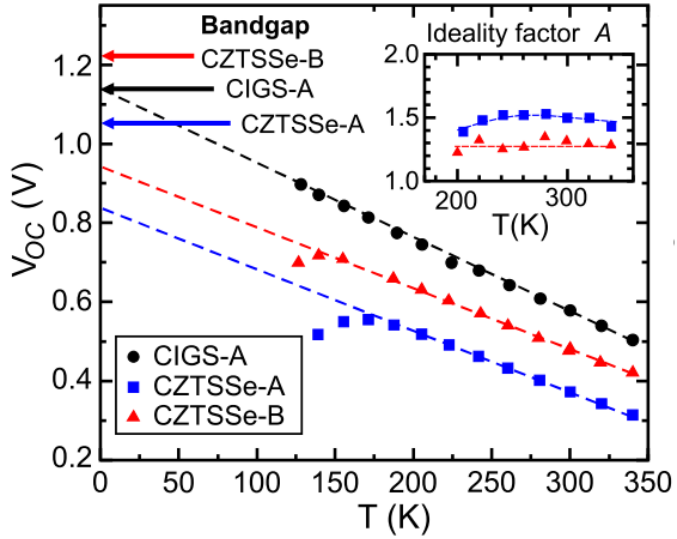

**Fig. 1.** (color online) Temperature dependence of $V_{OC}$ for CZTSSe and CIGSSe solar cells. Reproduced with permission from Ref. [10].

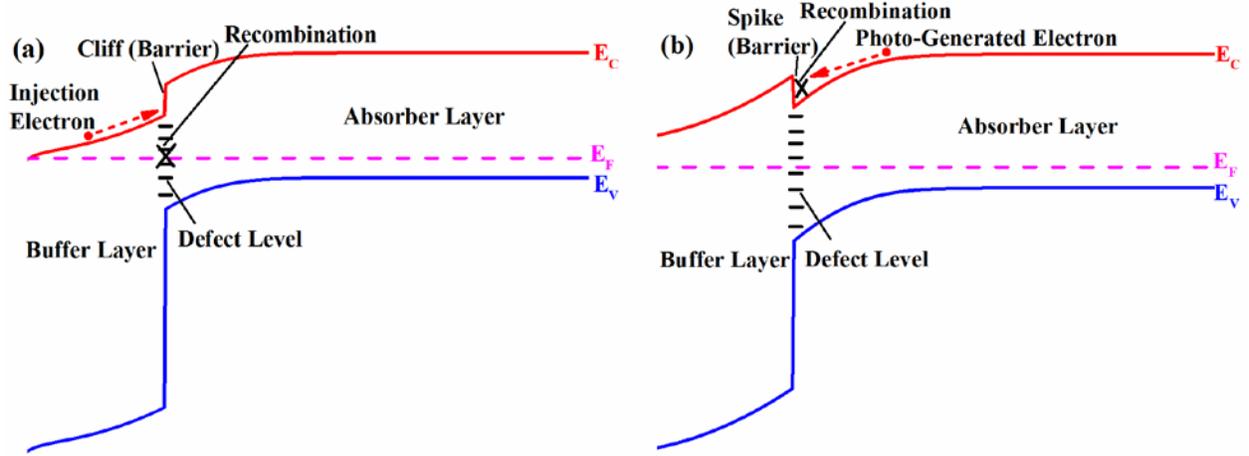

**Fig. 2.**(color online) Schema of band alignment at buffer/absorber interface: (a) cliff like, and (b) spike like, respectively. Spike and cliff are barriers for photo-generated and injection electrons, respectively. Reproduced with permission from Ref.[15].

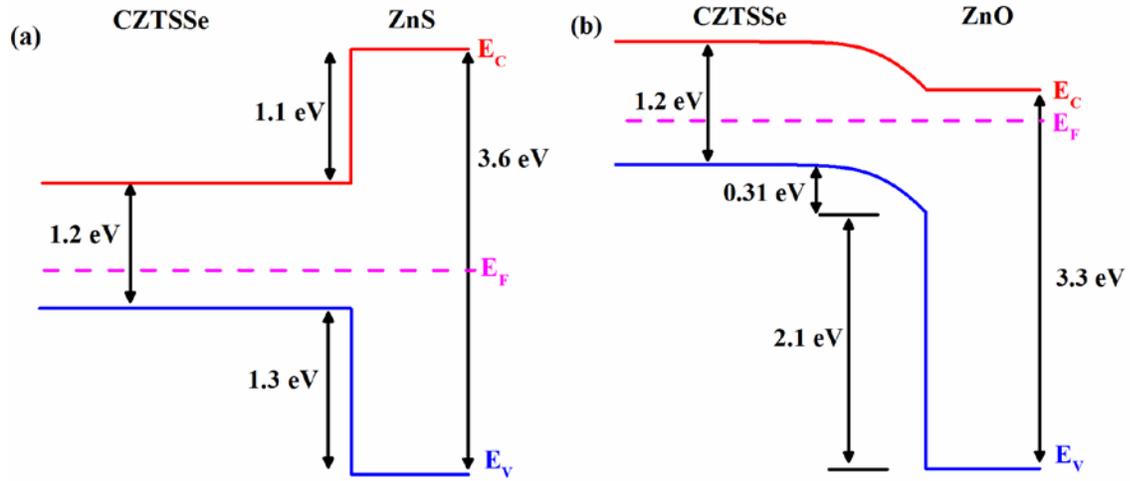

**Fig. 3.** (color online) Schematic of the band alignment between CZTSSe and (a) ZnS and (b) ZnO, respectively. Reproduced with permission from Ref. [30].

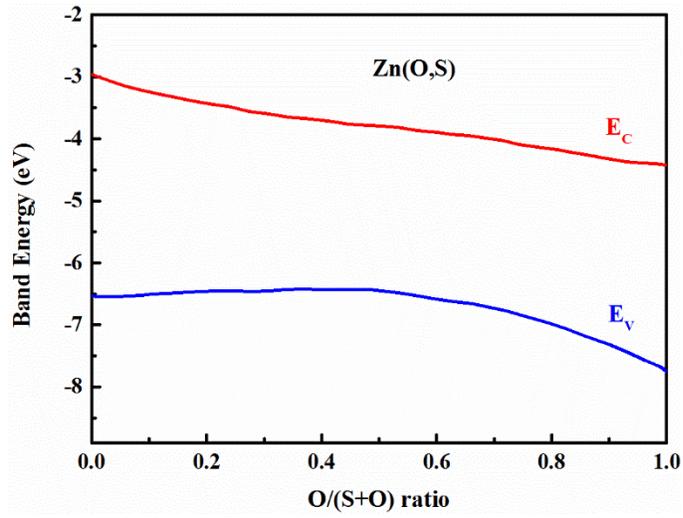

**Fig. 4.**(color online) Band energy diagram of Zn(O,S) as a function of O/(S+O). Reproduced with permission from Ref. [31-32].

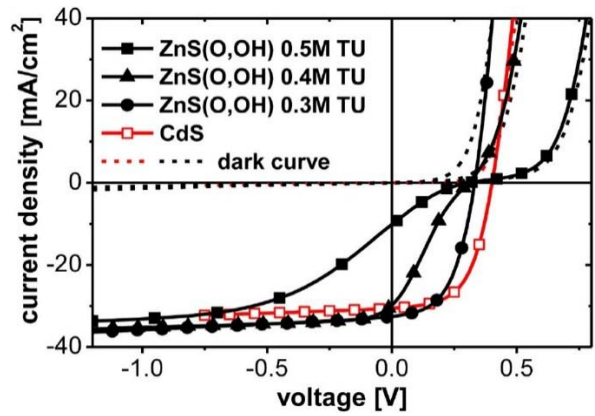

**Fig. 5.** (color online) Light and dark *J-V* curves of CZTSe solar cells with Zn(O,S) buffer layer prepared with different TU concentration during the CBD process and CdS as a reference buffer layer. Reproduced with permission from Ref. [33].

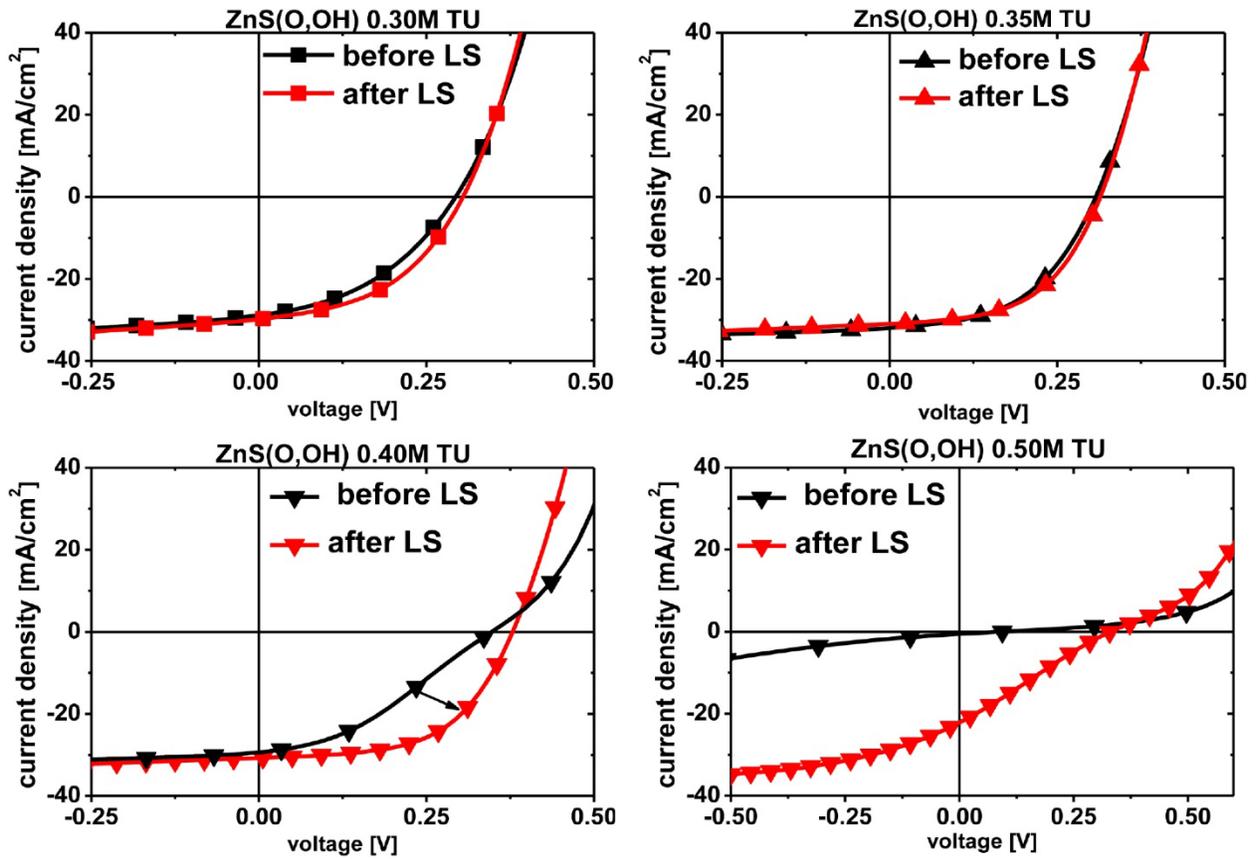

**Fig. 6.** (color online) Light *J-V* curves of CZTSe solar cells with Zn(O,S) buffer layer prepared with different TU concentration during the CBD process before and after light soaking for 260 min. Reproduced with permission from Ref. [33].

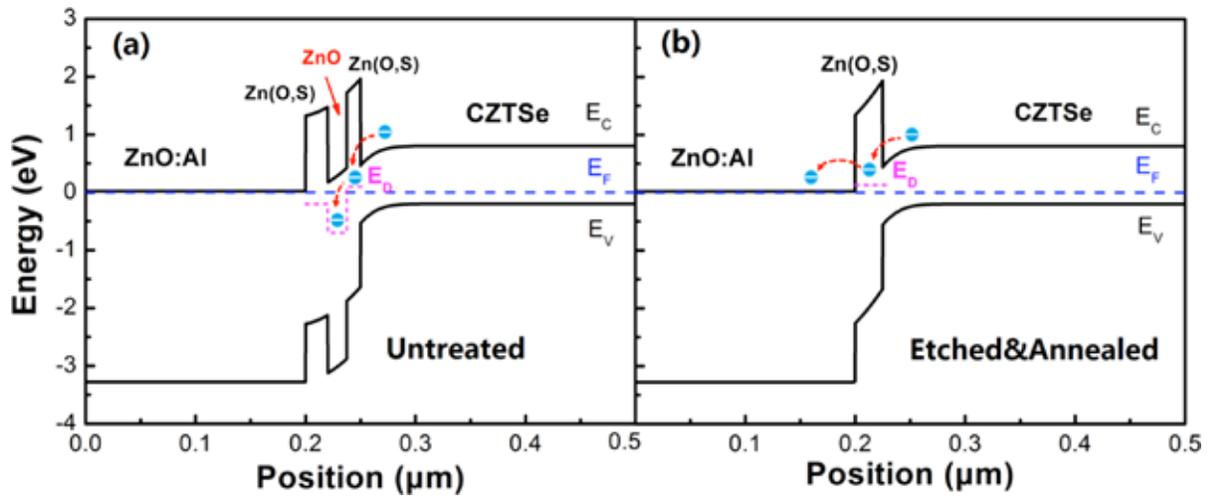

**Fig. 7.**(color online) The schematic diagrams of band alignments of the Zn(O,S)/CZTSe interface for (a) the Untreated sample and (b) the Etched & annealed sample. Reproduced with permission from Ref. [28].

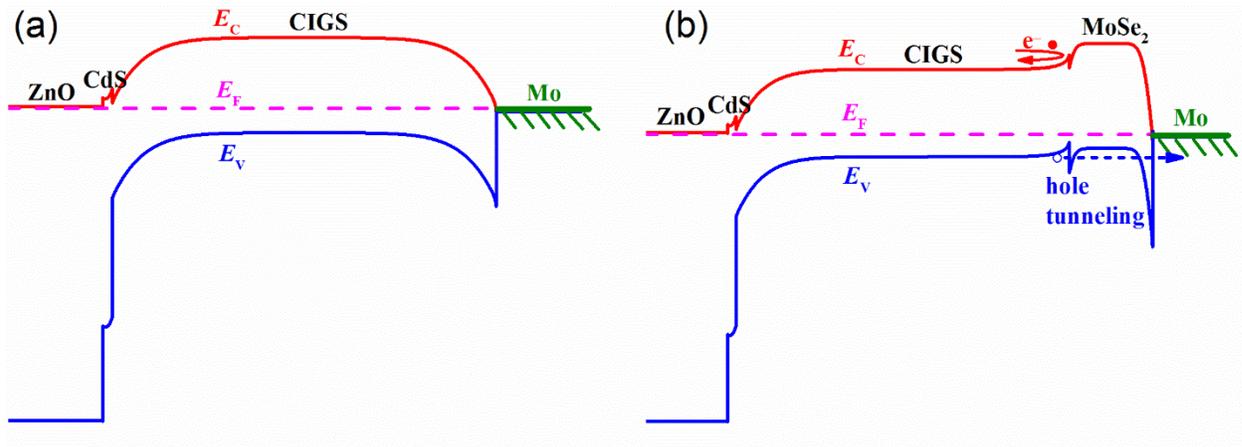

**Fig. 8.**(color online) Schematic diagram of the band structure of the CIGS solar cells: (a) without MoSe$_2$ layer, and (b) with MoSe$_2$ layer.

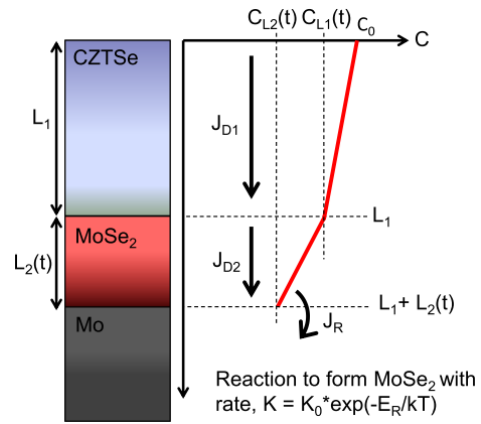

**Fig. 9.** (color online) Schematic of excess Se concentration profile across CZTSe/MoSe$_2$/Mo during annealing in a steady-state. Reproduced with permission from Ref. [69].

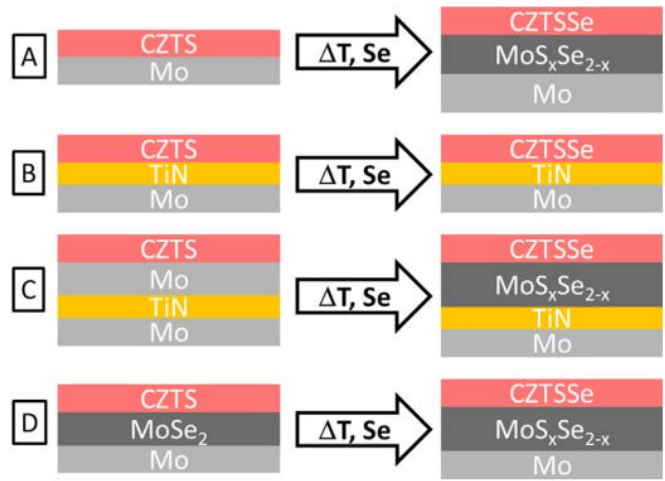

**Fig. 10.** (color online) Schematic drawings of the different back contact configurations. Reproduced with permission from Ref. [72].

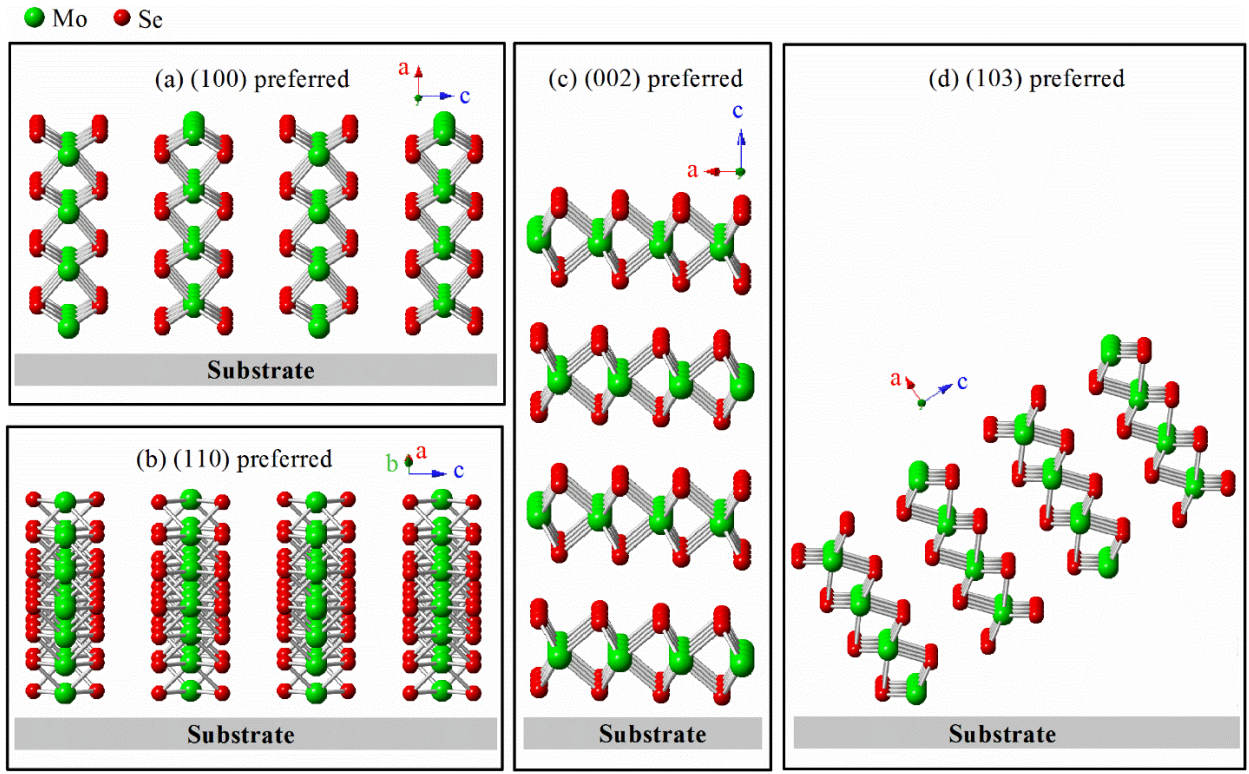

**Fig. 11.** (color online) Crystal structures of MoSe$_2$ along different direction, which is created with CrystalMaker software: (a) (100) preferred orientation with Se-Mo-Se sheets perpendicular to the substrate, (b) (110) preferred orientation with Se-Mo-Se sheets perpendicular to the substrate, (c) (002) preferred orientation with Se-Mo-Se sheets parallel to the substrate, and (d) (103) preferred orientation with Se-Mo-Se sheets tilted to the substrate. Reproduced with permission from Ref. [81].

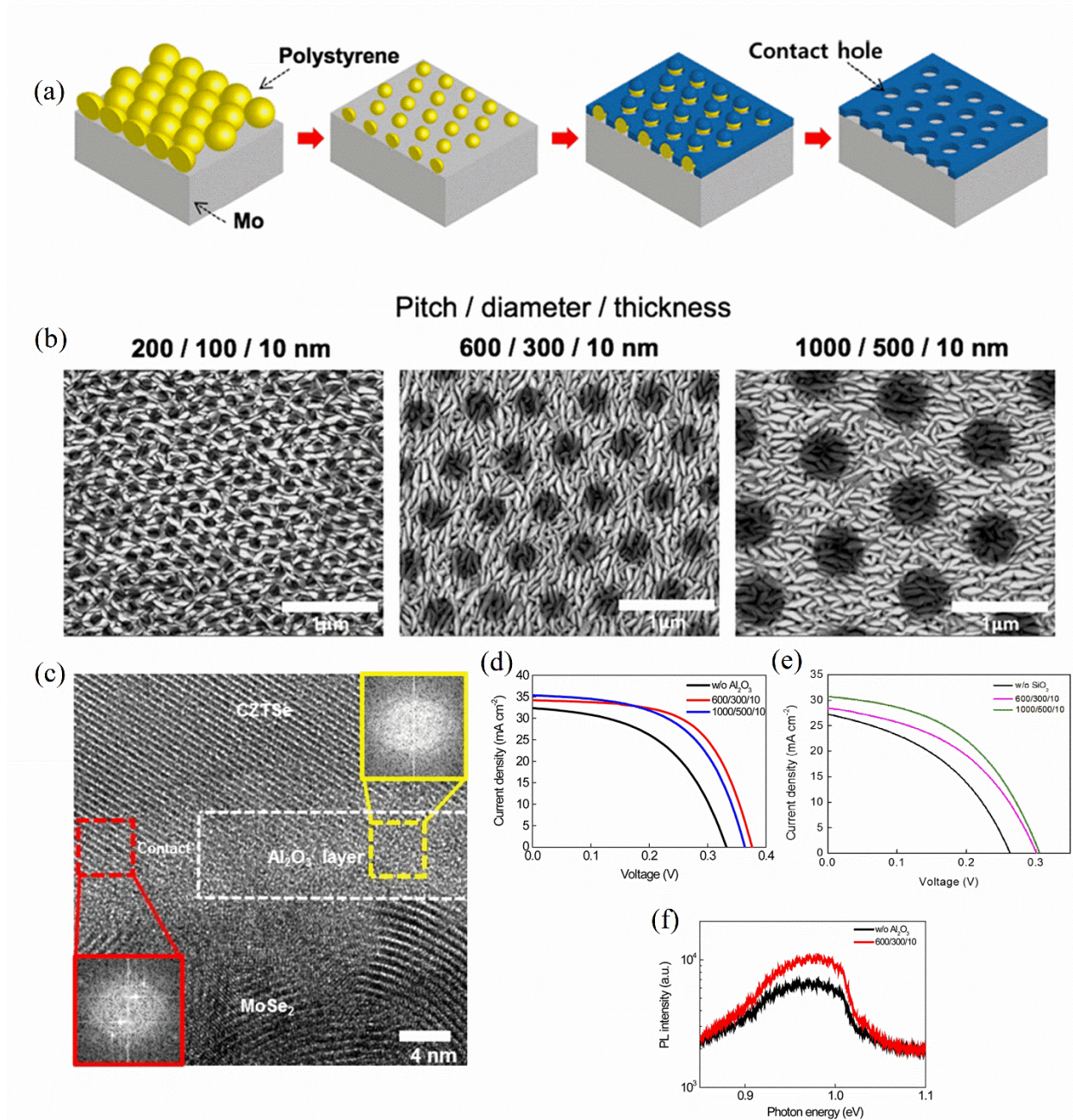

**Fig. 12.** (color online) (a) Schematic drawings of passivation layer with nano-patterned local contacts. (b) SEM images of contact holes on Mo after nanosphere lithography. (c) Cross-sectional TEM image of a bottom $Al_2O_3$ passivation layer (white dotted square) with a local contact between CZTSe and $MoSe_2$. Insets are Fourier transform images taken by the area of near the surface (yellow dotted square) and the bulk (red dotted square), respectively. (d) *J-V* characteristics for the CZTSe solar cells with and wihout a (d) $Al_2O_3$ and (e) $SiO_2$ passivation layer at the CZTSe/Mo interface. (f) 10 K photoluminescence spectra of CZTSe solar cells with and without an $Al_2O_3$ passivation layer at the CZTSe/Mo interface. Reproduced with permission from Ref. [88].

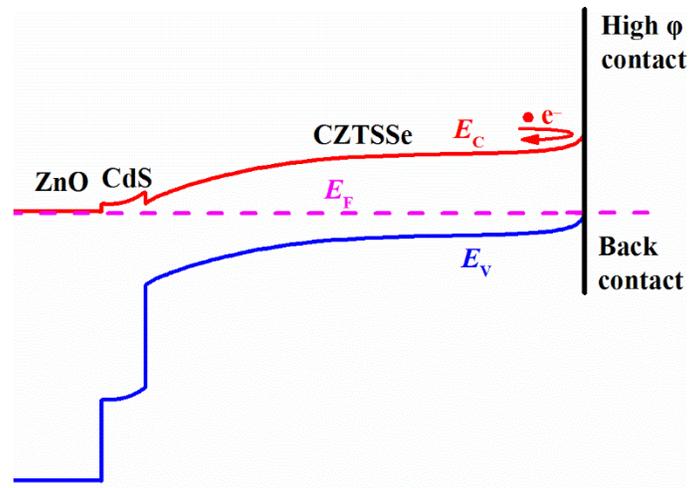

**Fig. 13.**(color online) Schematic band structure schematic of a CZTSSe solar cell with a high work function material between CZTSSe absorber and Mo back contact.

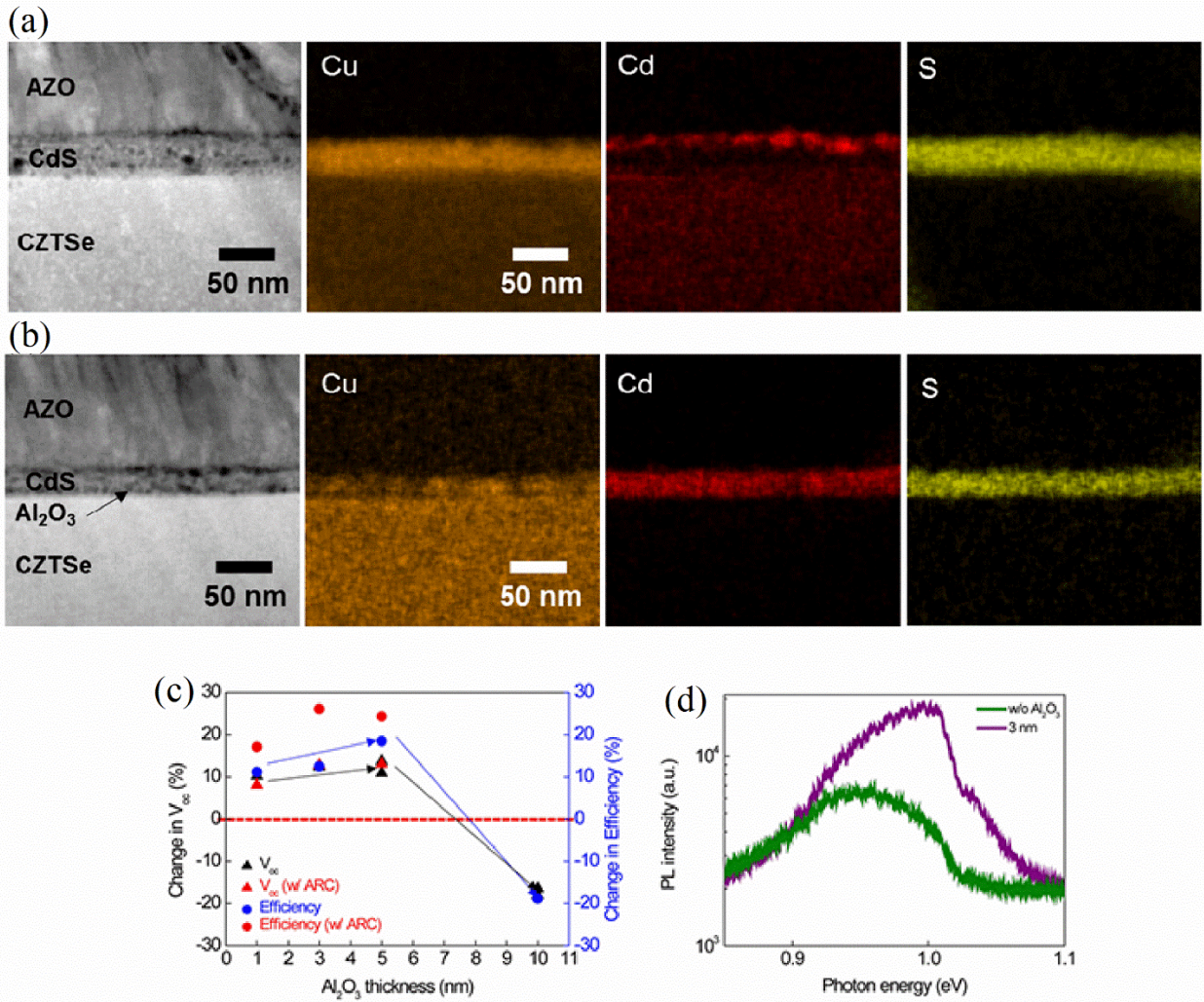

**Fig. 14.** (color online) HAADF-STEM images and corresponding EDS mapping of Cu, Cd, and S of CZTSe solar cells (a) without and (b) with an $Al_2O_3$ passivation layer at the CZTSe/CdS interface. (c) Changes in $V_{oc}$ and efficiency as a function of $Al_2O_3$ thickness (solid lines are guides to eyes and a dashed line indicates a respective reference). (d) 10 K PL spectra of CZTSe solar cells with (purple) and without (green) a top $Al_2O_3$ passivation layer at the CZTSe/CdS interface. Reproduced with permission from Ref. [88].

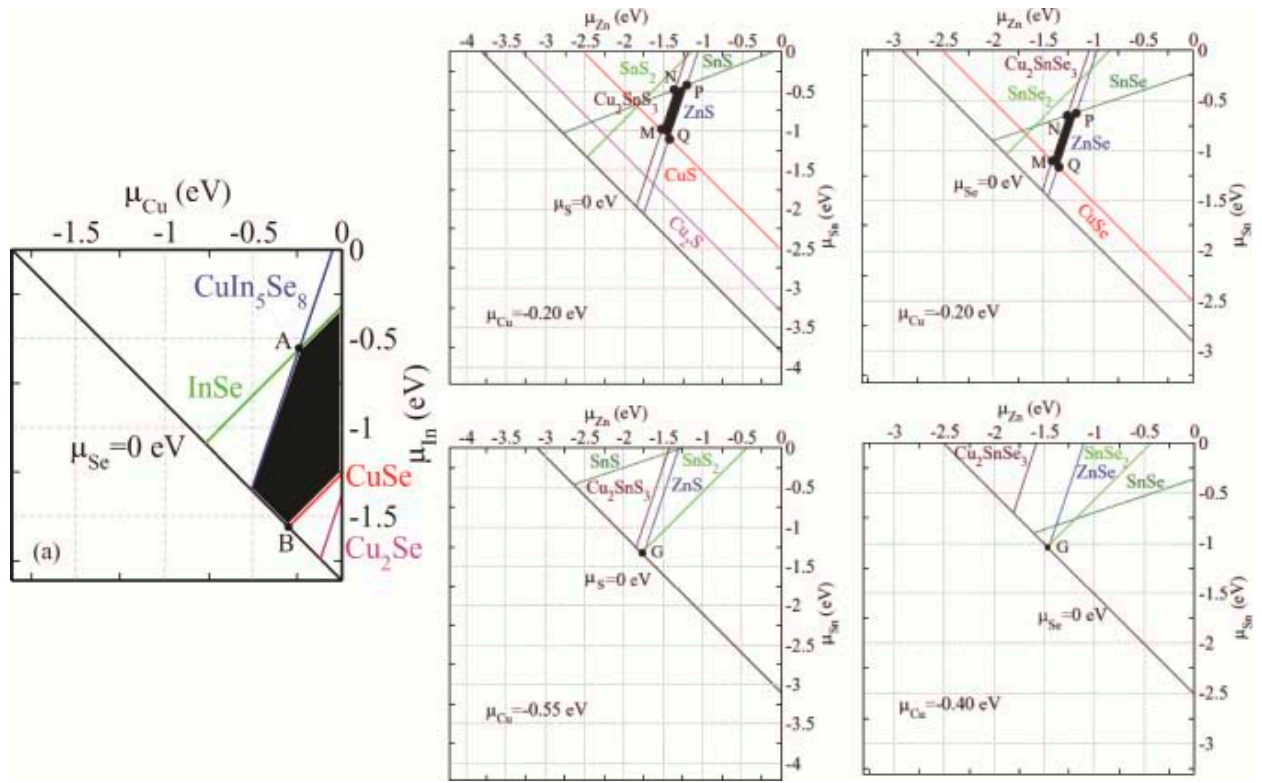

**Fig. 15.**(color online) The calculated phase stable region of CISe (left), CZTS (center), and CZTSe (right). Reproduced with permission from Ref.[102].

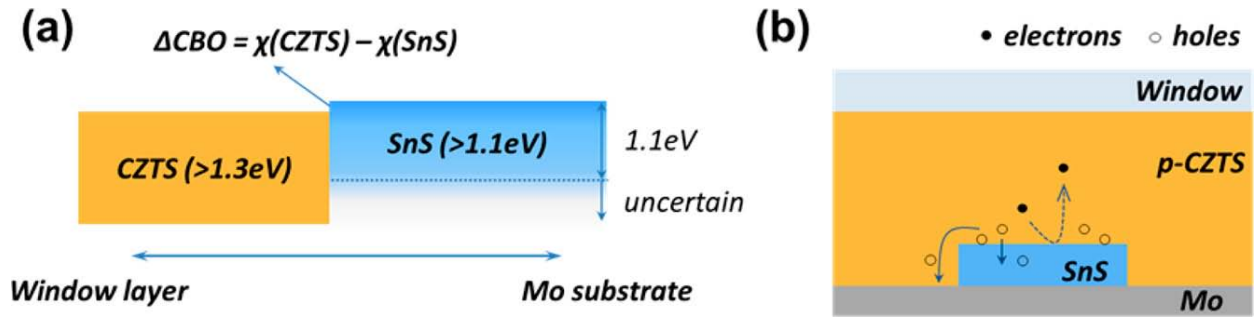

**Fig. 16.**(color online) (a) Band edges at the interface of CZTS/SnS, and (b) charge carrier transport paths close to CZTS/SnS interface. Reproduced with permission from Ref. [133].

**Table 1** A comparison of photovoltaic performance and $V_{OC}$ deficit for record CZTS(e)-based solar cells with their theoretical performance and with CIGS. SQL-CZTSSe refers to the theoretical performance of the CZTSSe solar cell.

|  | $E_g$ [eV] | $V_{OC}$ [mV] | $J_{SC}$ [mA/cm$^2$] | FF [%] | Eff. [%] | $V_{OC}$ deficit [mV] | Ref. |
|---|---|---|---|---|---|---|---|
| SQL-CZTSSe | 1.15 | 887 | 42 | 89 | 32.8 | 263 | [9] |
| CIGS | 1.10 | 741 | 37.8 | 80.6 | 22.6 | 359 | [1] |
| CZTSe | 1.01 | 423 | 40.6 | 67.3 | 11.6 | 578 | [135] |
| CZTSSe | 1.13 | 513.4 | 35.2 | 69.8 | 12.6 | 617 | [8] |
| CZTS | 1.50 | 700 | 21.3 | 63.0 | 9.4 | 800 | [136] |

**Table 2** Summary of conduction band offset (CBO) at CdS/CZTS or CdS/CZTSe interface, respectively.

| Interface | CBO [eV] | Ref. |
|---|---|---|
| CdS/CZTS | -0.33 | [16] |
| | 0.41 | [7] |
| | 0.0 | [17] |
| | -0.06 | [18] |
| | -0.3 | [20] |
| | -0.15 | [21] |
| CdS/CZTSe | 0.48 | [7] |
| | 0.34 | [19] |
| | 0.55 | [21] |